\documentclass{aa}
\usepackage{graphicx}
\usepackage{natbib}
\usepackage{txfonts}
\usepackage{color}
\bibpunct{(}{)}{;}{a}{}{,} 
\begin{document}
   \title{Radiance and Doppler shift distributions across the network of the quiet Sun}
   \author{H. Tian\inst{1,2}
           \and
           C.-Y. Tu\inst{1,2}
           \and
           L.-D. Xia\inst{3}
           \and
           J.-S. He\inst{2}
          }
   \offprints{H. Tian}
   \institute{Max-Planck-Institut f\"ur Sonnensystemforschung, Katlenburg-Lindau, Germany\\
              \email{tianhui924@gmail.com}
             \and
             Department of Geophysics, Peking University, Beijing, China\\
             \email{chuanyitu@pku.edu.cn}
             \and
             School of Space Science and Physics, Shandong Univ. at Weihai, Weihai, Shandong, China\\
             \email{xld@ustc.edu.cn}
             }
   \date{}
\abstract
{}
{The radiance and Doppler-shift distributions across the solar
network provide observational constraints of two-dimensional
modeling of transition-region emission and flows in coronal funnels.
These distributions have not, however, been studied in detail and we
attempt an investigation for a quiet Sun region.}
{Two different methods, dispersion plots and average-profile
studies, were applied to investigate these distributions for three
EUV lines. In the dispersion plots, we divided the entire quiet Sun
region scanned by SUMER into a bright and a dark part according to
an image of Fe~{\sc{xii}} taken by EIT during the scanning; we
plotted intensities and Doppler shifts in each bin as determined
according to a filtered intensity of Si~{\sc{ii}}. We also studied
the difference in height variations of the magnetic field as
extrapolated from the MDI magnetogram, in and outside network, in
the two parts. For the average-profile study, we selected 74
individual cases and derived the average profiles of intensities and
Doppler shifts across the network. Cases with large values of blue
shift of Ne~{\sc{viii}} were further studied. }
{The dispersion plots reveal that the intensities of Si~{\sc{ii}}
and C~{\sc{iv}} increase from network boundary to network center in
both the bright and dark parts. However, the intensity of
Ne~{\sc{viii}} shows different trends, namely increasing in the
bright part and decreasing in the dark part. In both parts, the
Doppler shift of C~{\sc{iv}} increases steadily from internetwork to
network center. The height variations in the magnetic field imply a
more homogeneous magnetic structure at greater heights and clearly
reflect the different magnetic structures in the two regions. The
average-profile study reveals that the intensities of the three
lines all decline from the network center to internetwork region.
The binned intensities of Si~{\sc{ii}} and Ne~{\sc{viii}} have a
good correlation. We also find that the large blue shift of
Ne~{\sc{viii}} does not coincide with large red shift of
C~{\sc{iv}}.}
{Our results suggest that the network structure is still prominent
at the layer where Ne~{\sc{viii}} is formed in the quiet Sun, and
that the magnetic structures expand more strongly in the dark part
than in the bright part of this quiet Sun region. Our results might
also hint for a scenario of magnetic reconnection between open
funnels and side loops. }

\titlerunning{Radiance and Doppler shift distributions across the network of the quiet Sun}
\authorrunning{H. Tian et al.}
\maketitle

\section{Introduction}
Our knowledge of the solar atmosphere, in particular the transition
region where temperature increases and density drops dramatically
with height, depends largely on spectroscopic observations.
Ultraviolet emission lines can provide ample information about the
physics of the upper chromosphere, transition region, and lower
corona. Since the plasma is optically thin for most emission lines,
we can extract information about the physical conditions prevailing
in their source regions from the line profiles (see reviews by
\cite{Mariska1992} and \cite{Xia2003}).

Before the launch of SOHO (the Solar and Heliospheric Observatory),
our understanding of the transition region were largely based on
observations made by early spectrographs such as the Naval Research
Laboratory (NRL) S082-B EUV spectrograph onboard the Skylab space
station \citep{BartoeEtal1977}, and NRL High-Resolution Telescope
Spectrograph (HRTS) flown on some rockets and Spacelab2
\citep{BruecknerEtal1977,BruecknerBartoe1983,BruecknerEtal1986}.
These early observations provided much valuable information about
the solar transition region, which was reviewed by
\cite{Mariska1992}.

Due to the high spectral and spatial resolution of the SUMER (Solar
Ultraviolet Measurements of Emitted Radiation) instrument
\citep{WilhelmEtal1995,LemaireEtal1997} onboard SOHO (Solar and
Heliospheric Observatory), many more line profiles were obtained,
identified and used in the studies
\citep{CurdtEtal2001,CurdtEtal2004}, and the Doppler shifts of these
EUV spectral lines can be measured with an accuracy of about
1-2~km/s \citep{BrekkeEtal1997,HasslerEtal1999,PeterJudge1999,
WilhelmEtal2000,PopescuEtal2004}. Besides, another spectrograph CDS
(Coronal Diagnostic Spectrometer) onboard SOHO also has the
capability of simultaneous multi-wavelength imaging in a number of
transition region and coronal lines \citep{HarrisonEtal1995}. Thus,
our understanding of the solar transition region was improved
significantly since 1995 when SOHO was launched.

One of the most prominent features in the chromosphere and
transition region is the network structure \citep{Reeves1976}, which
is the upward extension of the supergranular boundary at greater
heights, and is characterized by clusters of magnetic flux
concentrations \citep{PatsourakosEtal1999}. The network manifests
itself as bright lanes in the radiance images of chromospheric and
transition region lines. The network cell has a size of
20,000-30,000 km \citep{SimonLeighton1964}, and lasts for 20 to 50
hours \citep{SimonLeighton1964, Schrijver1997, RajuEtal1998}. The
height variation in the network size can be found in
\cite{PatsourakosEtal1999}, \cite{GontikakisEtal2003}, and
\cite{TianEtal2008b}.

\cite{Gabriel1976} proposed the first magnetic network model, in
which the transition region emission originates from funnels
diverging with height from the underlying supergranular boundary.
However, this model failed to reproduce the emission measure at
temperatures below $10^5 K$ \citep{Athay1982}. \cite{DowdyEtal1986}
discovered many fine-scale structures of mixed polarities in the
photospheric magnetic network; they proposed a modified model in
which only a fraction of the network flux opens into the corona as a
funnel shape, while the rest of the network is occupied by a
population of low-lying loops with lengths less than $10^4$ km. In
this model, the small loops are heated internally. Based on SUMER
observations, \cite{Peter2001} suggested a new sketch of the
structure of the solar atmosphere, which includes different
structures that dominate at different heights. The most intriguing
feature in this structure is that there are two types of coronal
funnels: funnels connected to the solar wind and funnels that are
feet of large loops. It might also be possible that the fundamental
structure of the solar transition region has not yet been resolved
by current observations \citep{Feldman1983, Feldman1987,
FeldmanLaming1994, DoschekEtal2004}.

Another prominent characteristic of the solar transition region is
the observed redshift of transition region lines (see a review in
\cite{Mariska1992}). The redshift can reach up to 15 km/s in the
network. In a statistical sense, these redshifts imply the presence
of plasma flows or wave motions in the quiet Sun with amplitudes
that are substantial fractions of the sound speed
\citep{PeterJudge1999}. Different mechanisms such as solar wind
outflows, spicules, siphon flow through loops, nano-flares,
explosive events, and downward propagating MHD waves have been
suggested to explain these prominent Doppler shifts of transition
region lines (see reviews in \cite{Mariska1992},
\cite{BrekkeEtal1997}, \cite{PeterJudge1999}, and \cite{Xia2003}).
The redshift is most significant in the middle transition region and
decreases to become a blue shift in the upper transition region. The
redshift to blueshift transition occurs at electron temperatures of
about $5\times10^5$ K and the Doppler shift variation with
temperature can be found in \cite{PeterJudge1999},
\cite{TeriacaEtal1999}, and \cite{XiaEtal2004}.

There is an increasing consensus that the observed large blue shifts
in the upper transition region and lower corona are associated with
solar wind outflows in coronal holes. Relative to past studies, a
far more accurate rest wavelength of Ne~{\sc{viii}}~($\lambda770$,
$2s^{2}S_{1/2}-2p^{2}P_{3/2}$) which originates in the upper
transition region and lower corona, were measured by
\cite{DammaschEtal1999}. The Doppler shift of this line can be
therefore determined well and has been intensively studied. Sizable
areas of large blue shift were found in the Ne~{\sc{viii}}
dopplergrams in coronal holes and considered to be a signature of
solar wind outflow \citep{HasslerEtal1999, Peter1999,
StuckiEtal2000, WilhelmEtal2000}; they tended to be larger in the
darker regions of coronal holes \citep{XiaEtal2003}.
\cite{HasslerEtal1999} studied the relationship between Doppler
shift and chromospheric network, and found that a larger blue shift
was closely associated with the underlying chromospheric network.
\cite{TuEtal2005a} inferred that the patches of large Ne~{\sc{viii}}
blue shift were associated with coronal funnels reconstructed by a
force-free field model.

In the quiet Sun, large Ne~{\sc{viii}} blue shifts were also found
in the network lanes, and considered to be a signature of the solar
wind \citep{HasslerEtal1999} or just mass supply to quiet coronal
loops \citep{HeEtal2007, TianEtal2008a}.

However, the distributions of intensities and Doppler shifts of EUV
lines across the network, which are required for 2-D modeling of
transition region emissions and flows in coronal funnels, have not
been investigated well. \cite{AiouazEtal2005b} developed a technique
of three-step filter, which was applied to an image of Lymann
continuum intensity to reveal the chromospheric network. They binned
the filtered continuum intensity and calculated the median value of
the Doppler shift of Ne~{\sc{viii}} in each intensity bin to produce
the so-called dispersion plot. They found that the maximum outflow
(blue shift) does not appear in the center of the network but rather
near network boundary. \cite{Aiouaz2008} extended this work by
studying the redshifts of transition region lines and found the
locations of the maximum inflows (redshift) of these lines are
network centers in the quiet Sun and network boundaries in the
coronal hole. \cite{XiaEtal2004} studied the correlation between
radiance and Doppler shift of Ne~{\sc{viii}} and found that a
slightly positive correlation exists. They also found that the
correlation is not clearly present for the C~{\sc{iv}} line.

Magnetic field extrapolation is a common way for the solar community
to study the magnetic coupling of different solar processes (see
e.g. \cite{WiegelmannNeukirch2002, RuanEtal2008}). With the
assumption that the magnetic field may be considered as almost
force-free at heights of about 400 km above the photosphere
\citep{MetcalfEtal1995}, \cite{MarschEtal2004} combined the
extrapolated magnetic field based on the force-free model proposed
by \cite{Seehafer1978} with EUV observations to study the plasma
flows in ARs and detected strong adjacent up and down flows
associated with sunspots. A similar method was applied successfully
by \cite{WiegelmannEtal2005}, \cite{TuEtal2005a},
\cite{TuEtal2005c}, and \cite{MarschEtal2006} for a coronal hole
study, in \cite{TuEtal2005b}, \cite{HeEtal2007}, and
\cite{TianEtal2008a} for a quiet sun study, and in
\cite{TianEtal2007} for the study of coronal bright points.

In this paper, we use two different methods, namely the method of
dispersion plot and average-profile study, to investigate the
distributions of radiance and Doppler shift of three EUV lines
across the network in a quiet Sun region.

\section{Observation and data reduction}
On 22 September 1996, a middle-latitude quiet-Sun region was scanned
by slit 2 ($1^{\prime\prime}\times300^{\prime\prime}$) of the SUMER
instrument from 00:40 to 08:15 UTC. The raster increment was about
$3^{\prime\prime}$. Three strong EUV lines, Si~{\sc{ii}} (153.3~nm),
C~{\sc{iv}} (154.8~nm), and Ne~{\sc{viii}} (77.0~nm) were included
in the selected spectral window. This data set was intensively
studied before by \cite{DammaschEtal1999}, \cite{HasslerEtal1999},
\cite{Peter2000}, \cite{GontikakisEtal2003}, \cite{TuEtal2005b}, and
\cite{TianEtal2008a}. More details about this observation can be
found in these papers. The selected area in our study is
$442^{\prime\prime}\times259^{\prime\prime}$ in size, which is the
same as that considered by \cite{TianEtal2008a}.

\begin{figure}
\resizebox{\hsize}{!}{\includegraphics{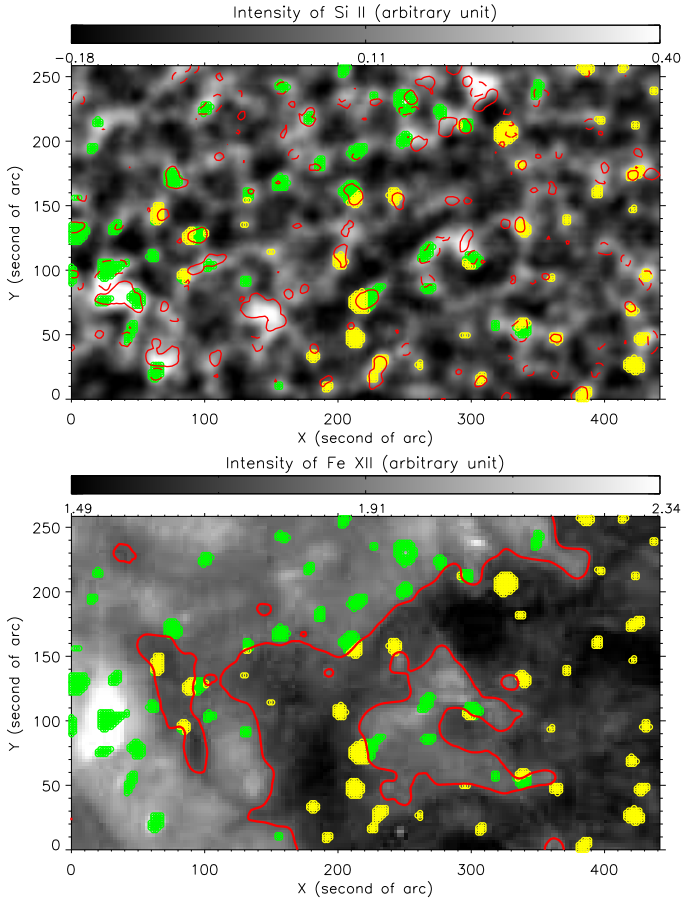}}
\caption{Upper: The intensity image of Si~{\sc{ii}}, the solid and
dashed contours represent strong magnetic fluxes with different
polarities (larger than 15 Gauss).  Lower: Image of
EIT-Fe~{\sc{xii}} observed at 02:30 on 22 September 1996, the red
lines mark the boundary between the bright part and the dark part.
The intensities are shown in a logarithmic scale. The green and
yellow circles mark positions of the data points in bins
representing the network center in Figs.~\ref{fig.3} and
~\ref{fig.4}, respectively.}\label{fig.1}
\end{figure}

The standard SUMER procedures for correcting and calibrating the
data were applied, including local-gain correction, flat-field
correction, geometrical-distortion, and dead-time correction.

Each observed spectrum was fitted by a single Gaussian curve,
complemented by a constant and a linear term describing the
background. By integrating over a fitted profile, we derived the
total count rate for the profile and thus obtained its intensity
that was then used to construct the intensity image of each line.
The Doppler shift of each line was determined by using the method
described by \cite{TuEtal2005b}.

We used the magnetogram taken by the Michelson Doppler Imager (MDI)
\citep{ScherrerEtal1995} onboard SOHO on the same day at 01:39 UTC.
The pixel size of the magnetogram is about $2^{\prime\prime}$. The
correction of the magnetogram and the coalignment of the magnetogram
with the SUMER images were carried out by applying the method
described by \cite{TuEtal2005b}. The linear Pearson correlation
coefficient between the coaligned magnetogram and Si~{\sc{ii}}
intensity image is 0.15 (the number of data points is 29436 and the
corresponding critical correlation coefficient is 0.01, so the
magnetic field and chromospheric emission correlate very well). The
Ne~{\sc{viii}} intensity image was used to coalign the SUMER and EIT
images. The uncertainties in both coalignments are less than
$2^{\prime\prime}$.

We normalized the intensities with respect to the median value over
the entire area for each line. The intensity image of Si~{\sc{ii}}
is presented in the upper panel of Fig.~\ref{fig.1}, where the solid
and dashed contours represent strong magnetic fluxes of different
polarities (higher than 15 Gauss). The network pattern is clear in
this chromospheric line and almost all the strong magnetic fluxes
are located in networks. The lower panel of the figure shows the
sub-region of a full-disk EIT(Extreme-Ultraviolet Imaging Telescope)
\citep{Delaboudini1995} image corresponding to the scan area of
SUMER. The EIT image was observed in Fe~{\sc{xii}} at 02:30 on 22
September 1996.

\section{The filtered intensity image of Si~{\sc{ii}}}

If the network pattern has a regular shape and the radiance
decreases from the center of the network to the internetwork, then
the radiance can be used to indicate the distance from the network
center. A strong, median, and weak radiance should be found close to
the network center, network boundary, and the internetwork region,
respectively. In Fig.~\ref{fig.1}, we can discern that the network
structure is very clear and thus the intensity of Si~{\sc{ii}} can
be used to represent the distance from the network center.

However, the intensity contrast between the network and internetwork
for different supergranular cell is different. To remove the local
brightenings and reduce the contrast, we applied a three-step filter
technique developed by \cite{AiouazEtal2005b} to the intensity image
of Si~{\sc{ii}}. The technique was applied instead to an intensity
image of Lymann continuum by \cite{AiouazEtal2005b}. The method can
be described in the following way: First we applied a median filter
to the image. The width of the median filter was chosen to be
30$^{\prime\prime}$, corresponding to the mean size of a
supergranular cell. The original intensity image was then divided by
the median-filtered image. Finally, the resulting image was smoothed
over 10$^{\prime\prime}$, which is the usually accepted value of the
network size at transition region temperatures
\citep{PatsourakosEtal1999}. By using this method, local
brightenings and intensity contrasts of network/internetwork for
different supergranular cells were reduced although remain in the
filtered image. We therefore applied two additional iterations of
this method: the method was reapplied to the image produced by the
previous filtering iteration. The final filtered intensity image of
Si~{\sc{ii}} is presented in Fig.~\ref{fig.7}.

In this filtered intensity image of Si~{\sc{ii}}, local brightenings
have been significantly reduced and the intensity contrasts between
the network and internetwork for different supergranular cells are
almost comparable. This filtered intensity was used to indicate the
distance from network center. A larger, median, and smaller value of
the filtered intensity of Si~{\sc{ii}} corresponds to a location in
the network center, network boundary, and internetwork region,
respectively. Then we are able to complete a statistic study on the
trend of a specific parameter from network center to internetwork
region, based on this conversion.

\section{Dispersion plots}

Our first method of studying the distributions of radiance and
Doppler shift across the network is the method of dispersion plot
previously employed by \cite{AiouazEtal2005b} and \cite{Aiouaz2008}.
We binned this filtered intensity image of Si~{\sc{ii}} and
calculated the average values of the original intensities and
Doppler shifts of the included lines in each bin to produce the
dispersion plots.

We divided the entire region into two parts of the same area
according to the intensity of EIT-Fe~{\sc{xii}}. The two parts are
labeled as bright part and dark part. The red lines in the lower
panel of Fig.~\ref{fig.1} indicate the boundary between the bright
and dark part. The Fig.3 of \cite{HeEtal2007} illustrated spatial
relations between intersection points of the extrapolated open field
lines at 4 Mm, which implies that the open field lines are located
in the dark part, if we compare it with Fig.~\ref{fig.1} of this
paper. While the bright part is occupied by a population of magnetic
loops of different heights \citep{TianEtal2008a}. Although the
correspondence between the open/closed field lines and the
dark/bright parts is not exact, a comparison between the statistical
study of the two parts should highlight a different property of the
two parts. Another reason why we divided the region into two parts
is that the network emission of Si~{\sc{ii}} in the bright part is
systematically higher than that in the dark part. It is therefore
more accurate to present dispersion plots for each part separately
to reduce the effects of mixing bright internetworks and dark
networks.

\begin{figure*}
\sidecaption
\includegraphics[width=13cm]{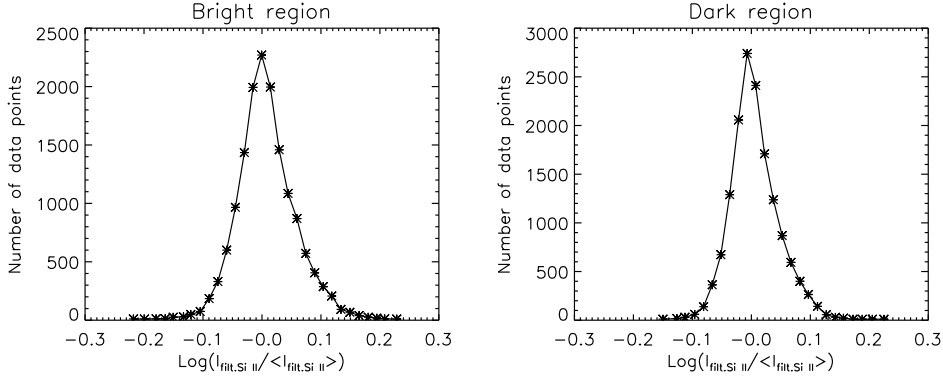}
\caption{Histogram of the filtered intensity of Si~{\sc{ii}} in the
bright part (left) and the dark part (right).} \label{fig.2}
\end{figure*}

To study the relationship between the filtered intensity of
Si~{\sc{ii}} (an indication of the distance from network center) and
the original radiance as well as Doppler shift, we first binned the
filtered intensity of Si~{\sc{ii}}. The number of data points in
each bin is shown in Fig.~\ref{fig.2}. The horizontal axis is given
by $log(I_{filt.Si~{\sc{II}}}/<I_{filt.Si~{\sc{II}}}>)$, where
$I_{filt.Si~{\sc{II}}}$ is the filtered intensity of Si~{\sc{ii}}
and $<I_{filt.Si~{\sc{II}}}>$ is the median value of the filtered
intensity of Si~{\sc{ii}} over the entire area. It appears that the
filtered Si~{\sc{ii}} intensity has a log-normal distribution in
both the bright and dark part. The sampling steps of the intensity
bins were approximately 0.015. The minimum number of data points per
bin was chosen to be $N_{min}=10$. If the number of data points in a
bin was less than 10, the step was changed to 0.030 to make a new
bin. If the number of data points in the new bin remained less than
10, we discarded this bin.

\begin{figure*}
\resizebox{\hsize}{!}{\includegraphics{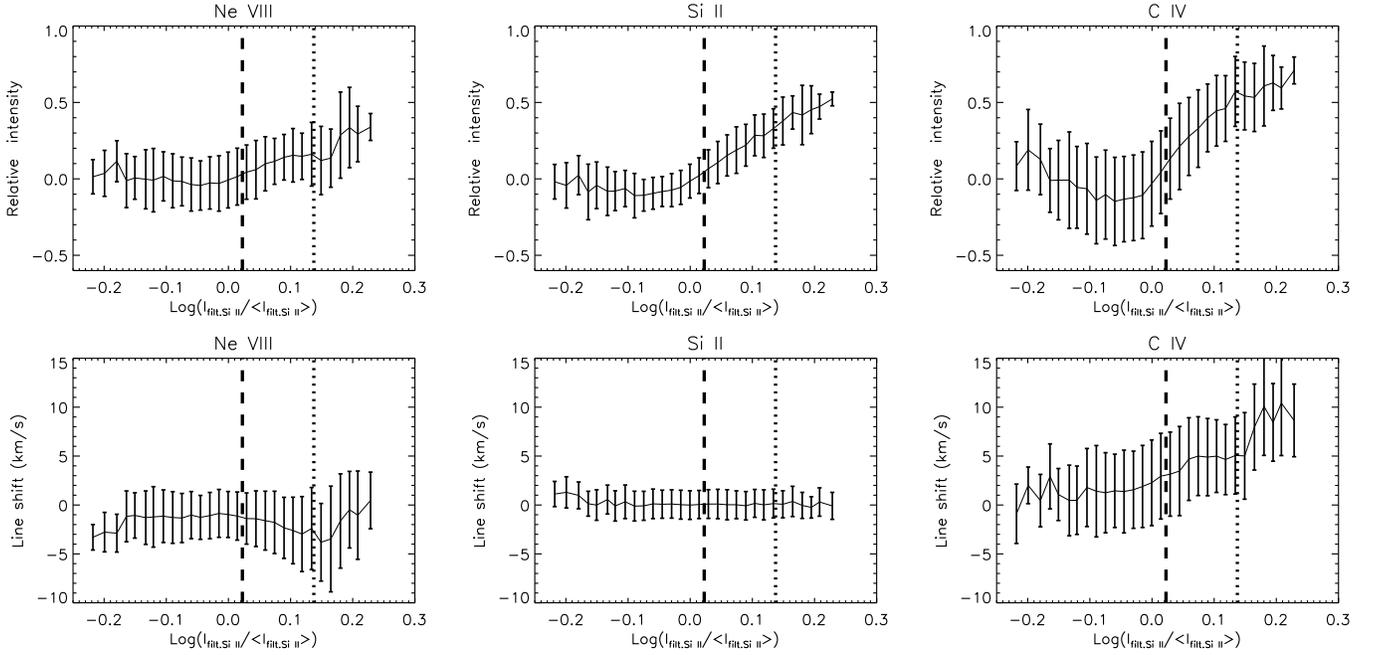}}
\caption{Dispersion plots for the bright part. The horizontal axis
is the filtered Si~{\sc{ii}} intensity ($log(I_{filt.Si~{\sc{II}}
}/<I_{filt.Si~{\sc{II}}}>)$), which can be considered to represent
the distance from network center. A higher $log(I_{filt.Si~{\sc{II}}
}/<I_{filt.Si~{\sc{II}}}>)$ represents a position closer to the
network center. The dashed line represents the outer edge of network
boundary and the dotted line marks boundary of network center.}
\label{fig.3}
\end{figure*}

\begin{figure*}
\resizebox{\hsize}{!}{\includegraphics{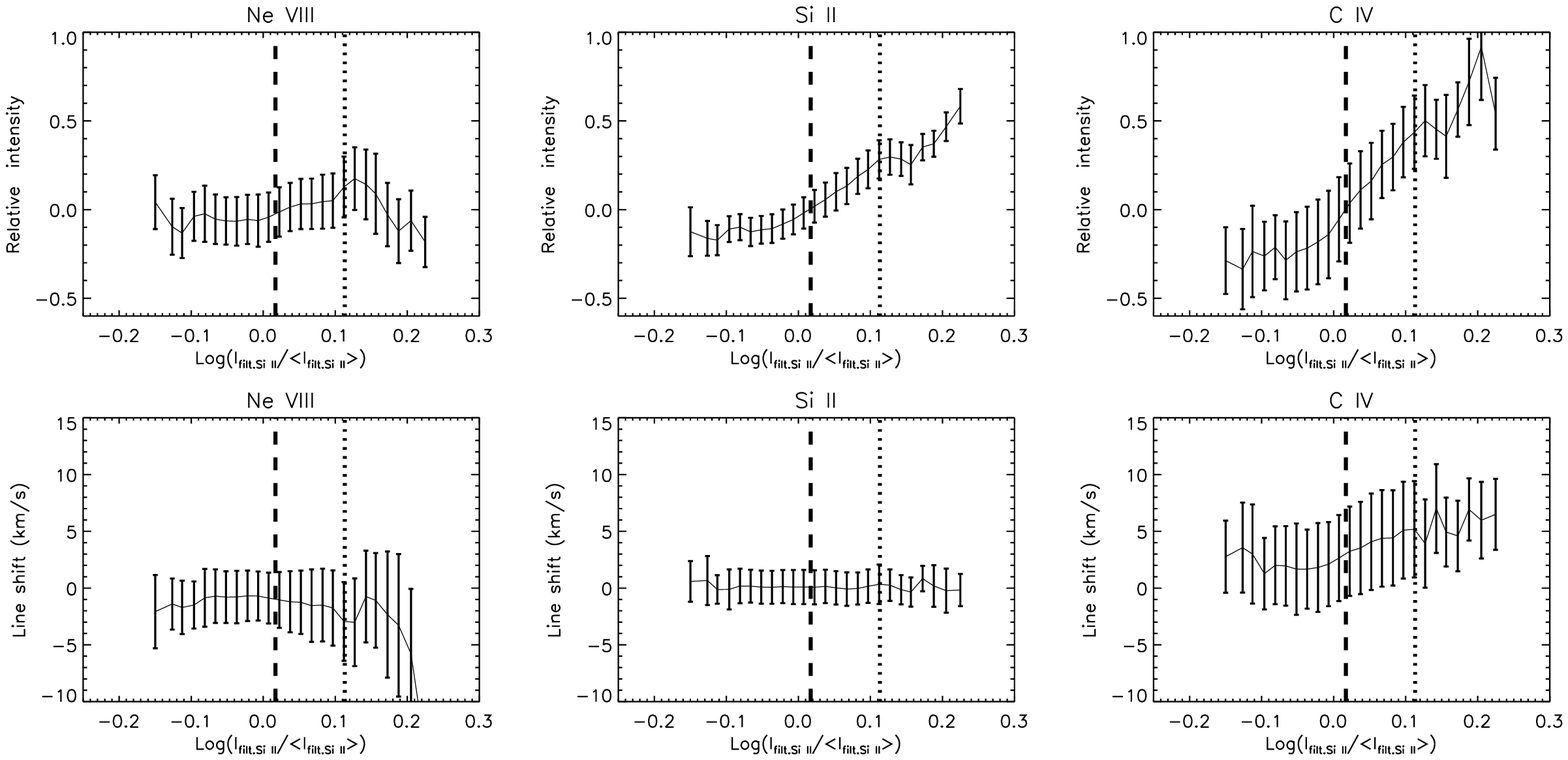}}
\caption{Dispersion plots for the dark part. The horizontal axis is
the filtered Si~{\sc{ii}} intensity
($log(I_{filt.Si~{\sc{II}}}/<I_{filt.Si~{\sc{II}}}>)$), which can be
considered to represent the distance from the network center. A
larger $log(I_{filt.Si~{\sc{II}} }/<I_{filt.Si~{\sc{II}}}>)$
represents a position closer to the network center. The dashed line
represents the outer edge of the network boundary and the dotted
line marks the boundary of the network center.} \label{fig.4}
\end{figure*}

\begin{figure*}
\sidecaption
\includegraphics[width=13cm]{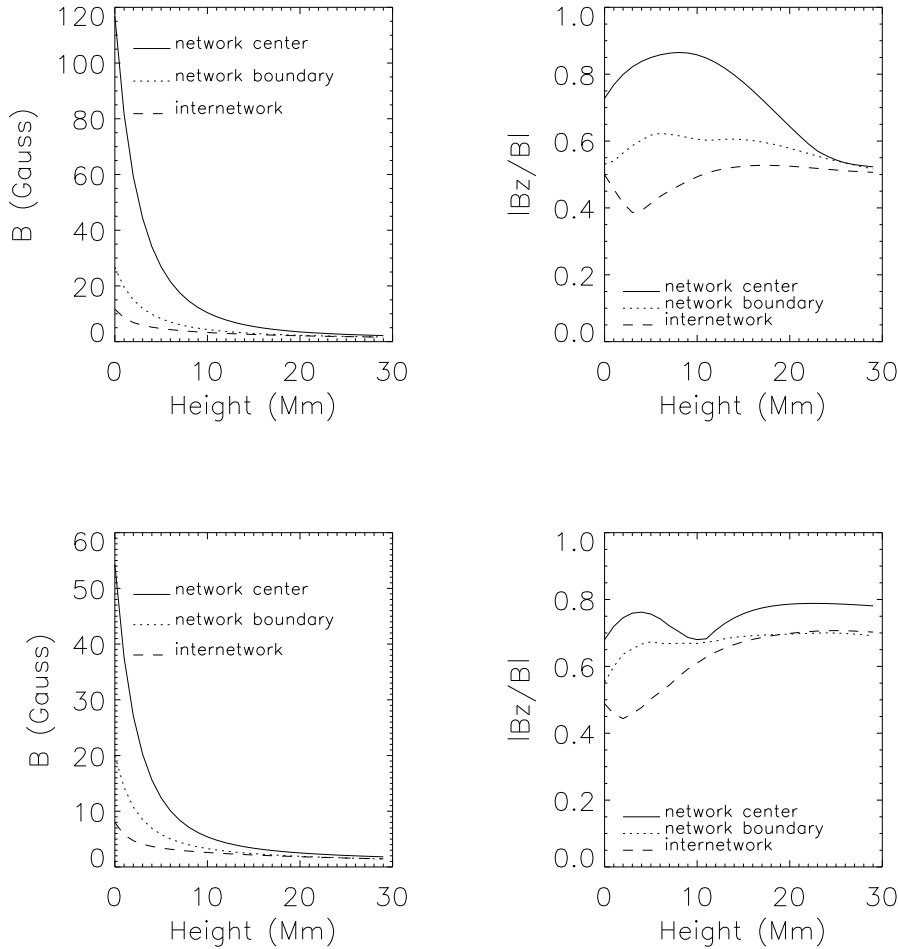}
\caption{The height variations in $B$ (left) and $\left\vert B_z/B
\right\vert$ (right) for the bright part.} \label{fig.5}
\end{figure*}

\begin{figure*}
\sidecaption
\includegraphics[width=13cm]{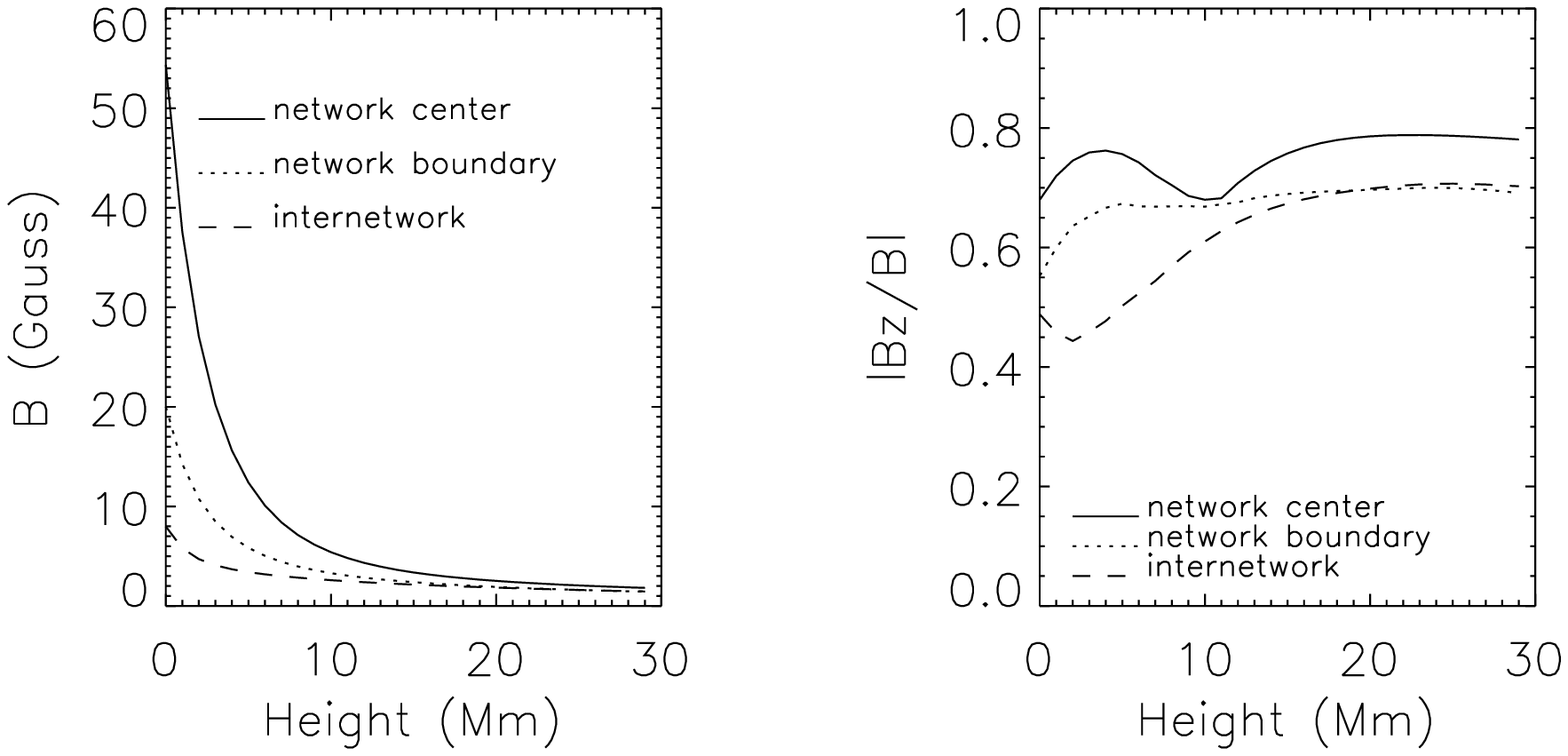}
\caption{The height variations in $B$ (left) and $\left\vert B_z/B
\right\vert$ (right) for the dark part.} \label{fig.6}
\end{figure*}

We then calculated the average values of the original intensities
and Doppler shifts of all three lines and the corresponding standard
deviations in each bin. The dispersion plots for the bright and dark
part are presented in Figs.~\ref{fig.3} and ~\ref{fig.4},
respectively. For each plot, the horizontal axis is the filtered
Si~{\sc{ii}} intensity ($log(I_{filt.Si~{\sc{II}}
}/<I_{filt.Si~{\sc{II}}}>)$), which might be considered to represent
the distance from the network center. The higher the filtered
Si~{\sc{ii}} intensity, the closer the distance to network center.
In each of the two parts, we can determine two thresholds for the
filtered Si ii intensity by assuming that the network occupies 1/3
of the entire area \citep{XiaEtal2004} and the network center
occupies 1/25 of the network area; these thresholds are indicated by
the dashed line representing the outer edge of the network boundary,
and the dotted line marking the boundary of the network center in
each panel of the figures. In each panel, we can consider that the
bins to the left of the dashed line are taken from the internetwork
(cell) region, the bins to the right of the dotted line are taken
from the network center, and the bins between the dashed line and
the dotted line represent the network boundary.

It is unsurprising that the intensities of Si~{\sc{ii}} and
C~{\sc{iv}} show a general increasing trend from network cell to
network center, since the network pattern is clearly discernible in
intensity images of chromospheric and transition-region lines. This
increasing trend is obvious within the network, indicating that
network emission increases steadily from the network boundary to
network center. The result might suggest that the magnetic fluxes
associated with the EUV emission within the network tend to cluster
towards the network center. By considering the error bars, the
chromospheric and transition-region emission appears, however, to be
distributed uniformly in the internetwork region, which suggests
that the associated internetwork magnetic fluxes are randomly
distributed.

Ne~{\sc{viii}} emission originates in the upper transition region
and lower corona. In Figs.~\ref{fig.3} and ~\ref{fig.4}, we find
that the intensity distribution of Ne~{\sc{viii}} between
internetwork and network center is similar to those of Si~{\sc{ii}}
and C~{\sc{iv}} in the bright part; in the dark part, however, they
are not consistent. From the network boundary to the network center,
the intensity of Ne~{\sc{viii}} shows different trends, increasing
in the bright part and decreasing in the dark part. Our previous
results demonstrated that magnetic structures expand throughout the
upper transition region and lower corona far more significantly with
height inside the coronal hole than the quiet Sun
\citep{TianEtal2008b}. And we also found that the dark part is
associated with open field lines as extrapolated from the
photospheric magnetogram \citep{HeEtal2007,TianEtal2008a}, which
suggests that the dark part has a similar characteristic (expanding
stronger than the normal quiet Sun) to that of the coronal hole.
Thus, we suggest the decrease of Ne~{\sc{viii}} intensity in the
network center is due to a more significant expansion of magnetic
structures in the dark part.

The Doppler shift of Ne~{\sc{viii}} also exhibits a different trends
between the network boundary and the network center in the two
parts, if we consider only the average value in each bin. In the
bright part, the maximum blue shift of Ne~{\sc{viii}} is close to
the network boundary, while it is found in the network center in the
dark part. We note that \cite{AiouazEtal2005b} claimed that the blue
shift of Ne~{\sc{viii}} is strongest at the edge of network patches
in a coronal hole, and the adjacent quiet Sun. They conjectured that
bipolar fields being swept in from the cell interior cause enhanced
magnetic activity with the more or less monopolar network fields and
lead to stronger shifts at the network boundary. The result for our
bright part data is similar to that of \cite{AiouazEtal2005b} and
the similar mechanism may also be responsible for it. But in the
dark part, where coronal funnels were found \citep{TianEtal2008a},
the strongest blue shift of Ne~{\sc{viii}} appears to be in the
center part of the funnel. By using different heating functions,
\cite{AiouazEtal2005a} developed a two-dimensional MHD model. For
one of the heating functions, the maximum blue-shift (outflow) was
found not to be in the very center of the funnel but in the vicinity
of the center, whereas, for the other heating function, the maximum
was aligned well with the center of the funnel. Considering this
model, the different locations at which Ne~{\sc{viii}} reaches its
maximum in the dark and bright part of our data set might be
explained as the result of different heating mechanisms dominating
in the two parts. However, the significant uncertainties in both our
measurements and those of \cite{AiouazEtal2005b} imply that the
above discussion is not very convincing.

\cite{Aiouaz2008} claimed that the maximum inflow (redshift) at
transition region temperatures is statistically toward the center of
the network in the quiet Sun and toward the boundary of the network
in the coronal hole, although large uncertainties were present in
their plots. In our case, the redshift of the transition region line
C~{\sc{iv}} increases steadily from internetwork to network center
in both the bright and dark parts, which is similar to the result
for the quiet Sun in \cite{Aiouaz2008}. The increase is more
pronounced in the bright than the dark part.

The Doppler shift of the chromospheric line Si~{\sc{ii}} line
remains at the zero level almost everywhere in and outside the
network, indicating that the structures of chromospheric loops are
similar everywhere in the lower solar atmosphere and there is a
balance between upflow and downflow in small chromospheric loops.

At the end of this section, we remind the reader of the principle
defect of this dispersion plot method. As stated above, the distance
to the network center is represented by, and the bins in
Figs.~\ref{fig.3} and ~\ref{fig.4} are divided according to, the
filtered intensity of Si~{\sc{ii}}. A comparable intensity contrast
between the network and internetwork for different supergranular
cells is required of a successful statistical study of the radiance
and Doppler shift distributions across the network. Although we
applied the three-step filtering method three times to reduce local
brightenings, there are still some relatively brighter and darker
networks in Fig.~\ref{fig.7}. In Fig.~\ref{fig.1}, the green and
yellow circles correspond to positions of the data points in bins
representing the network center in Figs.~\ref{fig.3} and
~\ref{fig.4}, respectively. It is clear that some data points at
network boundaries were classified into the network center bins. The
results in the network center bins were a mixture of contributions
from some network centers and boundaries. In contrast, there are
some networks that are insufficiently bright and have a lower
filtered intensity of Si~{\sc{ii}}. The intensities and Doppler
shifts in the centers of these networks were not considered as
contributions to those of the network center bins. This mixture
effect is difficult to remove. Applying the three-step filtering
method repeatedly can reduce the intensity contrast between the
network and internetwork for different supergranular cells. It can
also severely destroy the original network pattern. Our choice of
three iterations should therefore be appropriate. However, the
mixture effect is still existing and should not be neglected in our
case. The above discussion might also be the case in
\cite{AiouazEtal2005b} and \cite{Aiouaz2008}.

\section{Height variations of the extrapolated magnetic field}
In our previous works \citep{TuEtal2005b,HeEtal2007,TianEtal2008a},
we extrapolated the magnetic field from the photospheric magnetogram
based on a force-free model for this data set. Here we used this
previous result but studied the height variations in the
extrapolated magnetic field at different locations relative to the
network center.

For each of the two parts, we first selected three sets of grid
points, which correspond to three bins (one bin at the network
center, the other two bins at the network boundary and internetwork)
of the filtered Si~{\sc{ii}} intensity, at 0 Mm in our calculation
box. We then calculated the average values of the total magnetic
field magnitude $B$ and the ratio $\left\vert B_z/B \right\vert$,
where $B_z$ is the vertical component of $B$, in each set. After
that, we selected three new sets of grid points which have the same
x(east-west direction) and y(south-north direction) coordinates as
those of the original sets of grid points at different heights in
our calculation box. We then calculated the average values of $B$
and $\left\vert B_z/B \right\vert$ in each new set at differen
heights. The height variations in $B$ and $\left\vert B_z/B
\right\vert$ in each set are presented in Fig.~\ref{fig.5} for the
bright part and Fig.~\ref{fig.6} for the dark part.

The left panels of the two figures clearly show that the magnetic
field is far stronger in the network than the network boundary and
internetwork at 0 Mm, which is a reflection of the well-known fact
that magnetic flux tends to cluster in the network. We can see that
the magnetic field strength decreases with height almost
exponentially at each of the three selected locations. The
difference between the network and internetwork field also decreases
with height and reveals a more homogeneous magnetic structure in the
upper part of the solar atmosphere.

$\left\vert B_z/B \right\vert$ is an indicator of the inclination of
a magnetic field line with respect to the horizontal direction. The
right panels of the two figures reveal that $\left\vert B_z/B
\right\vert$ is much larger inside the network than at the network
boundary and internetwork region at lower heights above the
photosphere. This is easy to understand because there are more open
field lines and funnel structures originating in the network than
the internetwork. Above 10 Mm, the difference in $\left\vert B_z/B
\right\vert$ between the three locations appears, however, to
decrease with height. In the bright part, they are equal to each
other at about 30 Mm, where the value of $\left\vert B_z/B
\right\vert$ is about 0.5. This result suggests a more homogeneous
magnetic structure dominated by large loops with all possible
inclinations in the upper part of the solar atmosphere
\citep{Peter2001}. We found that the values of $\left\vert
B_z/B\right\vert$ at 30 Mm are far larger (about 0.7-0.8) than 0.5
in the dark part, which is consistent with our previous result that
there are open field lines \citep{HeEtal2007,TianEtal2008a} in this
part. We note that the value of $\left\vert B_z/B\right\vert$ at 30
Mm above the network center is larger than values at the other two
selected locations. This result may be explained that open field
lines tend to stretch up vertically from the network centers.

\section{Average-profile study}
The second method used in our study of the distributions of radiance
and Doppler shift across the network is an average-profile study.
This method aims to study the average trend of radiance and Doppler
shift from internetwork region to network by selecting some
individual cases.

\begin{figure}
\resizebox{\hsize}{!}{\includegraphics{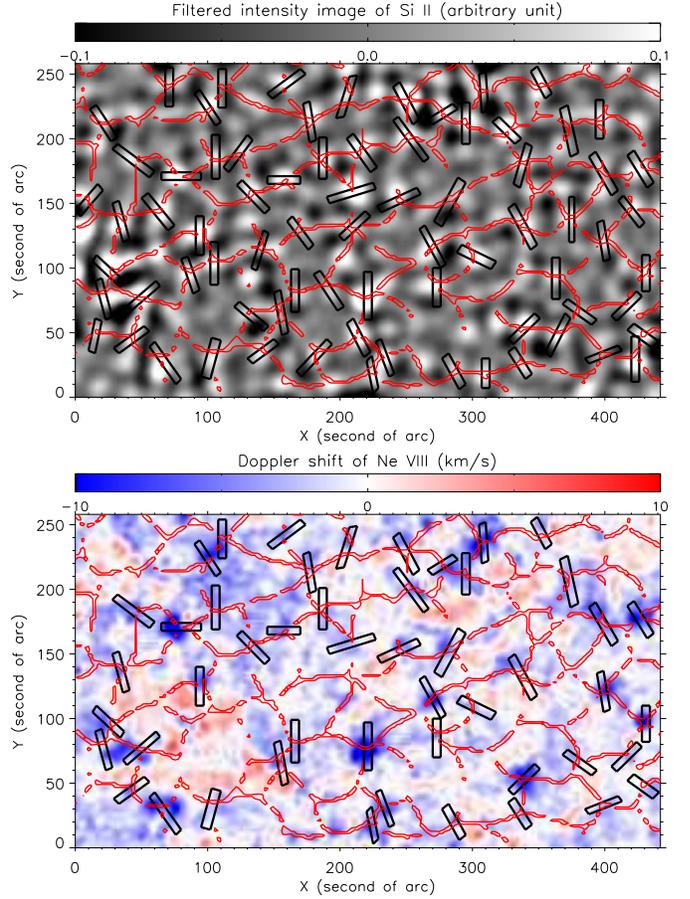}}
\caption{Upper: the filtered intensity image of Si~{\sc{ii}}. The
selected cases are outlined in black. Bottom: the Dopplergram of
Ne~{\sc{viii}}. The large-blueshift cases are outlined in black. In
both panels, the red network lines are taken from
\cite{HasslerEtal1999}.} \label{fig.7}
\end{figure}

We chose 74 parallelogram regions, each of which stretches from one
network cell to a neighboring cell across the bridging network; in
our case they are close to the places where more than two network
lanes intersect. In Fig.~\ref{fig.7}, the selected cases are
outlined in black in the filtered intensity image of Si~{\sc{ii}}.
In both panels, the red network lines are taken from
\cite{HasslerEtal1999} and can be used as a good approximate of the
network pattern.

\begin{figure*}
\resizebox{\hsize}{!}{\includegraphics{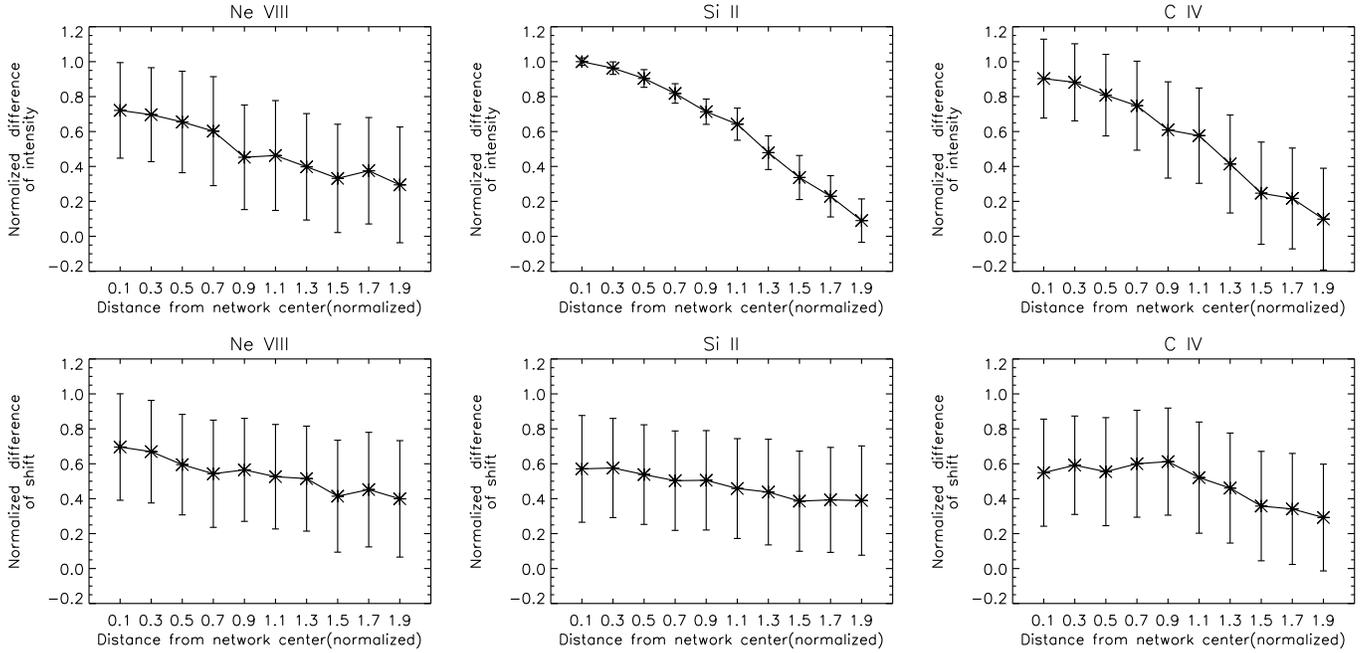}}
\caption{Average profiles of radiance and Doppler shift for the
selected cases.} \label{fig.8}
\end{figure*}

For each case, we first averaged the value of
$log(I_{filt.Si~{\sc{II}}}/<I_{filt.Si~{\sc{II}}}>)$ along the short
side of the parallelogram region and obtained the distribution of
$log(I_{filt.Si~{\sc{II}}}/<I_{filt.Si~{\sc{II}}}>)$ along the long
side. We found the position of the maximum value of
$log(I_{filt.Si~{\sc{II}}}/<I_{filt.Si~{\sc{II}}}>)$ in the
parallelogram region and defined this to be the network center, we
then defined the two locations where
$log(I_{filt.Si~{\sc{II}}}/<I_{filt.Si~{\sc{II}}}>)$ declines to
92\% of its maximum value to be the network boundaries (still in the
parallelogram region) on both sides of the network center. By using
this definition we can calculate the average size of the 74
networks. We obtained the value of $9.88\pm0.29$ arcsec, which is
consistent with the result of \cite{PatsourakosEtal1999}.

We then defined the network center to be the position $\emph{0}$,
and the two network boundaries to be position $\emph{1}$. We
extended the range $\emph{0-1}$ twice on both sides. The new
boundaries were defined as $\emph{2}$. We divided the range
$\emph{0-2}$ into 10 bins according to the position on each side,
that is $\emph{0-0.2}$, $\emph{0.2-0.4}$, $\emph{0.4-0.6}$,
$\emph{0.6-0.8}$, $\emph{0.8-1.0}$, $\emph{1.0-1.2}$,
$\emph{1.2-1.4}$, $\emph{1.4-1.6}$, $\emph{1.6-1.8}$,
$\emph{1.8-2}$, which we referred to as the normalized distance from
the network center. Then the two bins on different sides of the
network center with the same normalized distance from the network
center are combined into one bin. Therefore, there were only 10 bins
for each case and we calculated the average values of intensity and
Doppler shift for every line in each bin.

We applied the method mentioned above to each case and obtained the
profiles of the intensity and Doppler shift of every line across the
network. We then normalized the intensity and Doppler shift in the
form of $\frac{A-A_{min}}{A_{max}-A_{min}}$, where $A$ is the value
of line parameter(intensity or Doppler shift), and $A_{min}$ and
$A_{max}$ are the minimum and maximum value of the parameter in the
normalized distance range $\emph{0-2}$. We named this quantity the
normalized difference of intensity or shift. For the Doppler shift
of Ne~{\sc{viii}}, $A$ is used as its minus value in order to study
the blue shift of this line. We then obtained the normalized
profiles of the line parameters across the network for each case.
For all 74 cases, we extracted the 74 normalized values of each
parameter in the bin at the same normalized distance, and calculated
the mean value and the associated standard deviation. Then we
obtained an average normalized profile of each line parameter, which
is shown in Fig.~\ref{fig.8}.

General trends of the radiance and Doppler shift distributions
across the network are revealed in Fig.~\ref{fig.8}. We found that
the emissions of the three lines all declines from the network
center to internetwork region. The binned intensities of
Si~{\sc{ii}} and C~{\sc{iv}} decrease slowly from network center to
network boundary and more significantly in the internetwork region.
They have a correlation coefficient of 0.99. The correlation
coefficient between the binned intensities of Si~{\sc{ii}} and
Ne~{\sc{viii}} is 0.95, indicating that the network structure is
still prominent in the upper transition region of the quiet Sun.

Although with large uncertainty, the general trend of the
Ne~{\sc{viii}} blue shift decreases from the network center to the
internetwork region. The redshift of C~{\sc{iv}} is higher in the
network and continues to decrease beyond network boundary. As for
Si~{\sc{ii}}, its Doppler shift appears to be slightly higher in the
network than the internetwork, which may be due to the stronger
magnetic activity in the network. These Doppler-shift curves can
provide a measure of only the average trend and we are unable to
draw firm conclusions because of the significant size of the
uncertainties.

\section{Result of large-blueshift cases}

It has been found that large blue patches seen in the dopplergram of
Ne~{\sc{viii}} are associated with the chromospheric network
\citep{HasslerEtal1999}. Since a large blue shift of Ne~{\sc{viii}}
is considered to be a signature of the solar wind
\citep{HasslerEtal1999, TuEtal2005a} or mass supply to quiet coronal
loops \citep{HeEtal2007, TianEtal2008a}, it would be interesting to
study the distribution of the Ne~{\sc{viii}} blue shift in those
network regions around where large blue shift of Ne~{\sc{viii}} is
found.

\begin{figure*}
\resizebox{\hsize}{!}{\includegraphics{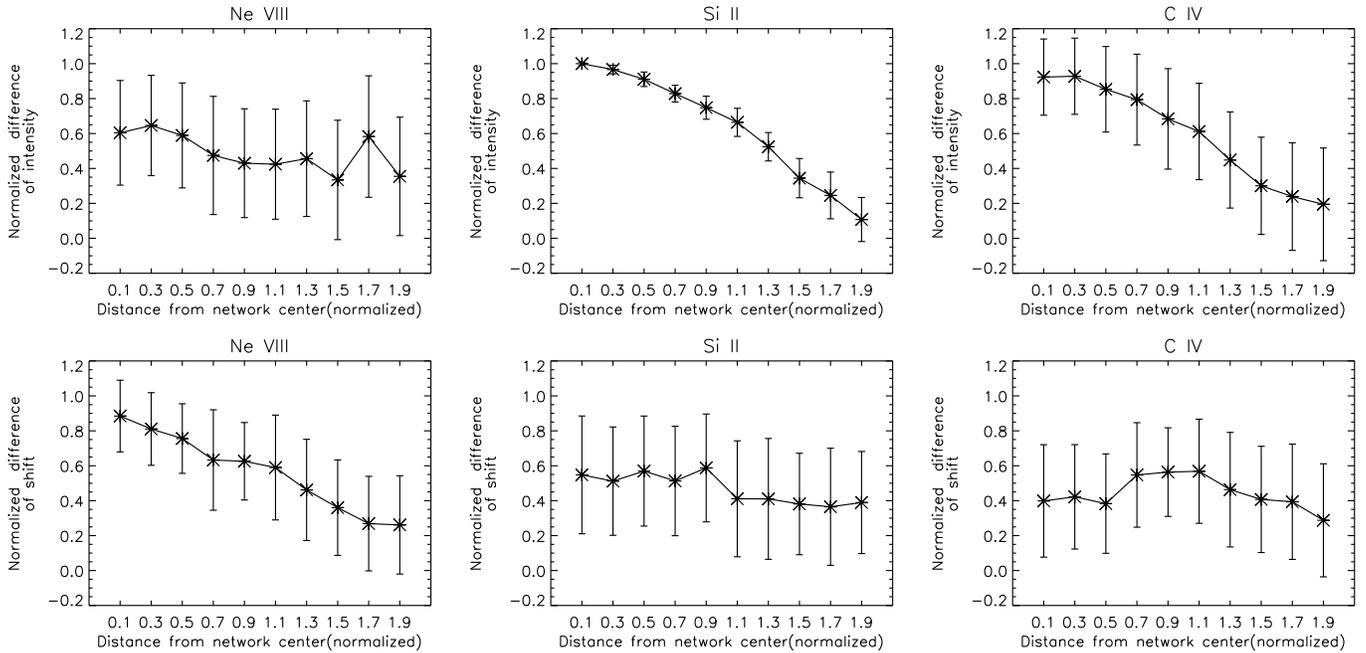}}
\caption{Average profiles of radiance and Doppler shift for the
large-blueshift cases.} \label{fig.9}
\end{figure*}

We selected cases for which the maximum Ne~{\sc{viii}} blue shift is
higher than 5 km/s in the normalized distance range $\emph{0-2}$,
from the 74 cases described in the previous section. We show 49
cases in the bottom panel of Fig.~\ref{fig.7}, for which our
criterion is met. The mean size of the network is $9.05\pm0.53$
arcsec for these large-blueshift cases. It is slightly smaller than
the average network size of all cases. We then constructed in a
similar way an average normalized profile of each line parameters.
The results are presented in Fig.~\ref{fig.9}.

The contrast between the intensity in and outside the network is
significant for Si~{\sc{ii}} and C~{\sc{iv}}, which is natural and
should be the case for almost every supergranular cell. However, the
intensity profile for Ne~{\sc{viii}} is rather flat, which suggests
that the magnetic funnels, whether being the footpoints of large
loops \citep{Peter2001} or associated with the solar wind
\citep{TuEtal2005a}, expand and stretch into the region above the
internetwork at the heights where considerable outflows of
Ne~{\sc{viii}} are found.

As for the Doppler shift, we found that the highest blue shift of
Ne~{\sc{viii}} does not coincide with the highest redshift of
C~{\sc{iv}}. The blue shift of Ne~{\sc{viii}} decreases steadily
from the network center to the internetwork region, while the
redshift of C~{\sc{iv}} reaches its maximum close to the network
boundary (the normalized distance $\emph{0.7-1.1}$). This result
appears to be consistent with the scenario proposed by
\cite{AxfordEtal1999} and \cite{TuEtal2005c}, in which reconnection
occurs between the open funnels and side loops. The outflows
produced by reconnections around a funnel tend to converge towards
the center of the funnel. In contrast, the hot plasma trapped in
low-lying loops are pulled down when they cool, and the downflows
are stronger at the boundary of the network where side loops are
accumulated. The Doppler shift of Si~{\sc{ii}} appears to be
slightly higher in the network than in the internetwork region. We
propose that it is due to the stronger magnetic activity in the
network, although the structures of chromospheric loops are similar
in and outside the network.

\

The detailed distributions of radiance and Doppler shift across the
network at different layers (chromosphere, transition region, and
corona) above the photosphere have not been studied systematically.
Our work with that of \cite{AiouazEtal2005b} and \cite{Aiouaz2008}
may be a first step in analyzing these distributions. The 1-D
modeling of the acceleration of the solar wind has been investigated
intensively. With the improvement of current observations, it is
increasingly important to develop 2-D and 3-D models to study the
origin and acceleration of the solar wind. We believe that the above
observation results are helpful for 2-D modeling of network emission
and outflows in coronal funnels.

\section{Summary}
We have completed a statistical analysis of the distributions of
both the radiance and Doppler shift across the network at different
layers (chromosphere, transition region, and lower corona) above the
photosphere. We applied the filtering technique developed by
\cite{AiouazEtal2005b} to an intensity image of the chromospheric
line Si~{\sc{ii}} obtained by SUMER/SOHO in a quiet Sun region. The
filtered intensity image of Si~{\sc{ii}} was used to reveal the
network. Two different methods, namely the method of dispersion plot
and the average-profile study, were used to investigate the
distributions of radiance and Doppler shift of Si~{\sc{ii}},
C~{\sc{iv}}, and Ne~{\sc{viii}} across the network.

We divided the entire quiet Sun region into a bright and a dark
part, and produced dispersion plots for each part. The intensities
of Si~{\sc{ii}} and C~{\sc{iv}} show an obvious increasing trend
from network boundary to network center in both parts. However, the
intensity of Ne~{\sc{viii}} shows different trends, namely
increasing in the bright part and decreasing in the dark part. This
result might suggest a stronger expansion of magnetic structures in
the dark part than in the bright part. In each of the two parts, the
Doppler shift of C~{\sc{iv}} increases almost steadily from
internetwork to network center. The average Doppler shift of the
chromospheric line Si~{\sc{ii}} remains at the zero level in and
outside network. The distribution of Ne~{\sc{viii}} Doppler shift is
not clear due to the large uncertainties. This method largely
depends on the intensity contrasts between the network and
internetwork for different supergranular cells. The mixture effect
influences the statistical results and could not be neglected
because it is difficult to obtain a comparable contrast for
different supergranular cells in real conditions.

The difference between height variations of the extrapolated
magnetic field above the internetwork, network boundary, and network
center was also investigated; this difference implies a more
homogenious magnetic structure at greater height and clearly
reflects the different magnetic structures in the two parts.

By selecting individual cases and investigating the average profiles
of radiance and Doppler shift distributions across network, we found
that the intensities of the three lines all decline from the network
center to internetwork region. The correlation coefficient between
the binned intensities of Si~{\sc{ii}} and Ne~{\sc{viii}} is 0.95,
indicating that the network structure is still prominent at the
layer where Ne~{\sc{viii}} forms in the quiet Sun. Although with
large uncertainties, the average Ne~{\sc{viii}} blue shift decreases
from the network center to internetwork region. The redshift of
C~{\sc{iv}} is higher in the network and decreases beyond the
network boundary.

Since the high blue shift of Ne~{\sc{viii}} could be a signature of
solar wind or mass supply to coronal loops, we also studied the
average profiles of radiance and Doppler shift across network for
the large-blueshift cases. The intensity distribution of
Ne~{\sc{viii}} does not correlate with those of Si~{\sc{ii}} and
C~{\sc{iv}}. We found that the blue shift of Ne~{\sc{viii}}
decreases steadily from the network center to the internetwork
region, and the large blue shift of Ne~{\sc{viii}} does not coincide
with large red shift of C~{\sc{iv}}.

Our observation results are worthy of consideration in terms of 2-D
modeling of network emissions and outflows in coronal funnels.

\begin{acknowledgements}
The SUMER project is financially supported by DLR, CNES, NASA, and
the ESA PRODEX programme (Swiss contribution). SUMER, MDI, and EIT
are instruments on board SOHO, an ESA and NASA mission. We thank the
teams of SUMER, MDI, and EIT for using their data. We also thank the
referee for his careful reading of the paper and for the comments
and suggestions.

H. Tian, C.-Y. Tu, and J.-S. He are supported by the National
Natural Science Foundation of China (NSFC) under contracts 40574078
and 40436015. H. Tian is now supported by China Scholarship Council
for his stay in the Max-Planck-Institut f\"ur Sonnensystemforschung
in Germany. L.-D. Xia is supported by NSFC under Grant 40574064 and
the Programme for New Century Excellent Talents in University
(NCET).

\end{acknowledgements}

\end{document}